\documentclass[pteplogo]{ptephy_v1}

\usepackage{graphicx}
\preprintnumber{01} 

\begin{document}
\title{First demonstration of gamma-ray imaging using balloon-borne emulsion telescope}


\author[1,*]{Hiroki Rokujo}
\affil{Nagoya University, Nagoya 464-8602, Japan\email{rokujo@flab.phys.nagoya-u.ac.jp}}
\author[2]{Shigeki Aoki}
\affil{Kobe University, Kobe 657-8501, Japan}

\author[3]{Kaname Hamada\thanks{Present Address: Nobeyama Radio Observatory, National Astronomical Observatory of Japan, Minamisaku 384-1305, Japan}}
\affil{Institute of Space and Astronautical Science, Japan Aerospace Exploration Agency (ISAS/JAXA), Sagamihara 252-5210, Japan}
\author[2]{Toshio Hara}
\author[2]{Tatsuki Inoue}
\author[1]{Katsumi Ishiguro\thanks{Present Address: Archaeological Institute of Kashihara, Kashihara 634-0065, Japan}}
\author[4]{Atsushi Iyono}
\affil{Okayama University of Science, Okayama 700-0005, Japan}
\author[1]{Hiroaki Kawahara}
\author{Koichi Kodama}
\affil{Aichi University of Education, Kariya 448-8542, Japan}
\author[1]{Ryosuke Komatani}
\author[1]{Masahiro Komatsu}
\author[2]{Tetsuya Kosaka}
\author[2]{Fukashi Mizutani}
\author[1]{Motoaki Miyanishi}
\author[1]{Kunihiro Morishima}
\author[1]{Misaki Morishita}
\author[1]{Mitsuhiro Nakamura}
\author[1]{Toshiyuki Nakano}
\author[1]{Akira Nishio}
\author[1]{Kimio Niwa}
\author[1]{Naoto Otsuka}
\author[2]{Keita Ozaki}
\author[1]{Osamu Sato}
\author[2]{Emi Shibayama}
\author[2]{Satoru Takahashi}
\author[2]{Atsumu Suzuki}
\author[2]{Ryo Tanaka}
\author[2]{Yurie Tateishi}
\author[2]{Shuichi Tawa}
\author[2]{Misato Yabu}
\author[2]{Kyohei Yamada}
\author[4]{Saya Yamamoto}
\author[1]{Masahiro Yoshimoto\thanks{Present Address: Gifu University, Gifu 501-1193, Japan}}


\begin{abstract}%
We promote the precise gamma-ray observation project Gamma-Ray Astro-Imager with Nuclear Emulsion (GRAINE), which uses balloon-borne emulsion gamma-ray telescopes.
The emulsion telescope realizes observations with high angular resolution, polarization sensitivity, and large aperture area in the 0.01--100 GeV energy region.
Herein, we report the data analysis of emulsion tracks and the first demonstration of gamma-ray imaging via an emulsion telescope by using the flight data from the balloon experiment performed in 2015 (GRAINE 2015).
The emulsion films were scanned by the latest read-out system for a total area of 41 m$^2$ in three months,
and then the gamma-ray event selection was automatically processed.
Millions of electron-pair events are accumulated in the balloon-borne emulsion telescope.
The emulsion telescope detected signals from a calibration source  (gamma rays from the interaction of cosmic rays with an aluminum plate) with a high significance during the balloon observation and created a gamma-ray image consistent with the source size and the expected angular resolution in the energy range of 100--300  MeV.
The flight performance obtained in the GRAINE 2015 experiment proves that balloon-borne emulsion telescope experiments with larger area are feasible while maintaining expected imaging performance.
\end{abstract}

\subjectindex{F10, H22}

\maketitle

\section{Introduction}

The experimental observation of cosmic gamma-ray emissions from black holes, pulsars, supernova remnants (SNRs), etc. is crucial to understand such high-energy objects and phenomena.
The Astro-Rivelatore Gamma a Immagini Leggero (AGILE) \cite{AGILE}, which was launched in 2007, and the Large Area Telescope on Fermi Gamma-ray Space Telescope (Fermi-LAT) \cite{LAT}, which was launched in 2008, have surveyed the sub-GeV/GeV gamma-ray sky.
Both experiments have achieved good results and contributed to the development of the gamma-ray astronomy, including the detection of more than 3000 gamma-ray sources \cite{catalog} and the discovery of cosmic-ray proton acceleration in SNRs \cite{cr,crfermi}. \par
Recently, new issues have come to light.
The observation at low galactic latitudes remains difficult because of the high density of gamma-ray sources and the increase in the background flux originating from galactic-diffuse gamma rays. Many gamma-ray sources remain unassociated on the galactic plane due to source confusions \cite{catalog}. Furthermore, unexpected gamma-ray excess in the galactic center region was reported \cite{GeVexcess}, which may arise from unsolved gamma-ray sources.
\par

Polarimetry data of high-energy photons are important to study the mechanism of the gamma-ray emission. 
The observation of polarization characteristics in the 100 MeV region, for example, is effective to investigate the curvature radiation from the outer gap region of pulsars \cite{J.Takata}.
Moreover, polarimetry of high-energy photons from gamma-ray bursts (GRBs) and active galactic nuclei (AGN) enable to perform new physics searches beyond the Planck scale.
Several observations in the hard X-ray and soft gamma-ray regions have been reported \cite{INTEGRAL, INTEGRAL2, GAP, PoGo}; however, to date, no substantial sensitive observations in the high-energy gamma-ray region ($>$10 MeV) have been performed.
In order to make progress in the gamma-ray astronomy, it is required not only a quantitative increase in the data observation, but also a qualitative improvement in their analysis.  \par
The Gamma-Ray Astro-Imager with Nuclear Emulsion (GRAINE) project aims at precisely observing gamma-ray sources using a new balloon-borne gamma-ray telescope. 
The high-angular resolution gamma-ray telescope, which is called emulsion telescope, consists of nuclear emulsion films. 
The nuclear emulsion technology allows a high-precision tracking of charged particles; in more detail, the incident angle and the position of the particle are measured with milliradian and submicron resolution in a thin material ($\sim10^{-3}$ in units of radiation-length, $X_0$).
After the exposure and the chemical development of the emulsion, it is crucial to perform a read-out process by using a microscope, because of the integrated detector.
Automated and fast scanning systems have been developed and practically used in accelerator neutrino experiments \cite{nakano}. 
\par
Since the nuclear emulsion technology allows to determine the incident angles of electrons and positrons at the vertex of pair-production processes ($\gamma + (\gamma) \rightarrow$ e$^{+}+ $e$^{-}$), the angular resolution for gamma rays (0.01--100 GeV) is approximately one order of magnitude higher than that of Fermi-LAT.
Additionally, the nuclear emulsion allows to measure the azimuthal angle of the plane where electron and positron tracks lie; thus, it has the sensitivity to the polarization of gamma rays \cite{polarization}.  
Nuclear emulsions do not have time resolution; hence, the gamma-ray incident time is provided by a new technique developed for the GRAINE project: the emulsion multi-stage shifter \cite{shifter}.
The time resolution has reached a subsecond accuracy, which is enough to be sensitive to the changes in the attitudes of a balloon-borne telescope with accuracy below one milliradian (the average rotation speed of the balloon is approximately 1 mrad/s). Therefore, it can point toward the gamma-ray directions determined by emulsion telescopes in celestial coordinates. 
\par
The emulsion telescope does not consist of electronic detectors; thus,
it is easy to increase its aperture area without deteriorating the resolution, which is typically caused by the limitation on the number of read-out electronic channels.
To perform our observations, we developed a large-area telescope with an aperture of about 10 m$^2$ and promoted long-duration balloon flights ($\sim$200 h), similarly to the JACEE or RUNJOB balloon-borne experiments  \cite{JACEE,RUNJOB}.
The target objects are sources lying on the galactic plane/center, bright and extended sources (for example, the SNR W44), galactic pulsars, GRBs, AGN, etc. \cite{COSPAR}.  
\par
In the first balloon experiment, the GRAINE 2011, the technical feasibility was demonstrated by means of a small-scale emulsion telescope and a daytime star camera system \cite{Rokujo,GRAINE2011}:
electron pair events were systematically detected and analyzed using emulsion track data;
the multi-stage shifter, which was introduced in the balloon experiment for the first time, provided the time information of the tracks recorded in emulsion films;
gamma-ray events plotted on the celestial coordinates with the incident angle, timing, and the attitude of the telescope.
\par
In the current phase of the GRAINE experiment, the goal is demonstrating the flight performance of an enlarged emulsion telescope. In particular, we focus on verifying the imaging performance (angular resolution) by detecting external gamma-ray sources (e.g. materials emitting gamma rays after having interacted with cosmic rays at balloon altitudes, and bright celestial gamma-ray objects).
For a significant detection, it is necessary to increase the number of detected gamma rays; to cope with this aim, since GRAINE 2011, we improved several aspects of the telescope and the emulsion scanning system \cite{GRAINE2015}. 
First of all, we improved the quality of the nuclear emulsion film. 
In GRAINE 2011, a hundred films with an area of 125-cm$^2$ were employed for the converter, the first part of the gamma-ray detection technology. 
However, the number of detected events was only 153, because the performance of the emulsion film was insufficient for an automatic offline process, and the analysis of the gamma rays was partially limited. 
To solve this issue, we adopted a new type of nuclear emulsion film with high efficiency and a good signal-to-noise (S/N) ratio (in the following we refer to such technology as ``highly sensitive emulsion films'').
The second improvement was the enlargement of the aperture area. We developed a multi-stage shifter that has a stage 30 times larger than that of the GRAINE 2011 model; in addition, we introduced a new pressure vessel in order to keep the large films vacuum-packed and flat. Furthermore, we developed a new scanning system for a faster data acquisition of the emulsion tracks.\par
After the implementation of the improvements specified above, we launched the second balloon experiment, the GRAINE 2015, on May 12, 2015. 
The balloon was launched from the Alice Springs balloon-launching station in Australia.
The balloon left the ground at 6:33 Australian Central Standard Time, and  reached an altitude of 37.2 km at 8:50.
The total flight duration was 14.4 h, including 11.5 h of level flight at an altitude of 36.0--37.4 km, and at a residual atmospheric pressure of 4.7--3.8 hPa.
After the balloon released the payload, the gondola landed approximately 130 km north of Longreach at 20:55.
On the next morning, all the payloads were recovered.\par
In this paper, we report the results of the analysis of the data recorded by the converter of the GRAINE 2015 emulsion telescope. 
(Detailed reports on the multi-stage shifter, the pressure vessel gondola, and the star camera system are in \cite{Mizutani, gondola, Camera}.)
We briefly introduce the detectors employed in the GRAINE 2015 experiment (section \ref{detectors-in-GRAINE2015}); then, we present the data recorded by the balloon-borne films and the quality obtained by using an emulsion scanning system (section \ref{data-taking}). In section  \ref{Analysis-for-selection}, we describe the automatic gamma-ray event selection using the flight data. Finally, in section \ref{demonstration-of-imaging}, we demonstrate the gamma-ray imaging using a calibration source during the balloon flight. 

\section{Experimental apparatus}\label{detectors-in-GRAINE2015}
Here, we summarize the detector technologies employed in the GRAINE 2015 experiment. More detailed information can be found in \cite{GRAINE2015}. 
\par
\begin{figure}
\begin{center}
\includegraphics[bb = 0 60 850 680, width=14cm,clip]{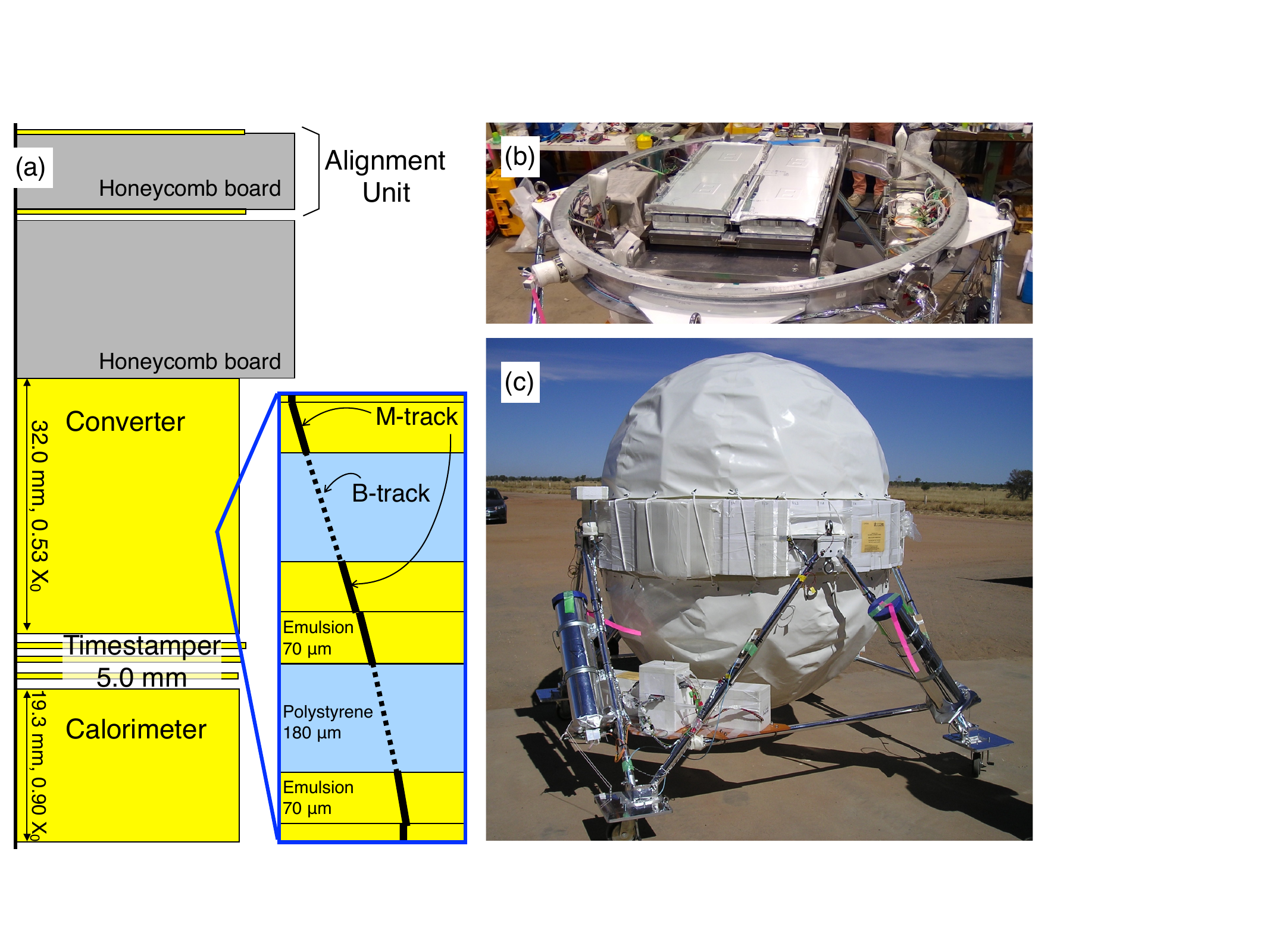}
  \end{center}
 \caption{
 (a) Cross-sectional scheme of emulsion films and the emulsion gamma-ray telescope. The telescope consists of alignment unit, converter, time stamper, and calorimeter.
 (b) Inside view of the pressure vessel. The emulsion gamma-ray telescope was placed at the center of a ring. The total aperture area of the four chambers was 3780 cm$^2$. 
 (c) Picture of the gondola. The spherical container (inside diameter of 1.6 m) is the balloon-style pressure vessel. The three-star cameras used as attitude monitors are deployed at the truss structure (one of them was on the other side).  
}
\label{detector}       
\end{figure}

Figure \ref{detector} shows  a cross-sectional scheme and a photograph of the emulsion gamma-ray telescope. The structure of the telescope consists of an alignment unit, a converter, a time-stamper, and a calorimeter, as schematically shown in Fig. \ref{detector}(a). 
The converter consists of 100 highly sensitive emulsion films. 
The emulsion films were produced in Nagoya University.
In order to develop a highly sensitive emulsion film, we increased the AgBr occupancy in the gelatin from the typical 30--45\% in volume to 55\%. In this way, 
the grain density (defined as the average number per unit length of the silver particles observed after the chemical development) was about 1.7 times higher than in typical films \cite{ozaki}.
The thicknesses of plastic and the emulsion layers on the both sides of the film are 180 $\mu$m and 70 $\mu$m, respectively.
The total thickness is 32 mm, which corresponds to 0.53 $X_0$.
The 34\% of vertical incident gamma rays that are converted into electron-positron pairs are recorded in emulsion films. 
The area of each converter film is 37.8 $\times$ 25 cm$^2$, and four units of the same structure were employed as shown in Fig. \ref{detector}(b), 
for a total aperture area of 3780 cm$^2$.\par
An alignment unit, which consists of two or three emulsion films kept vacuum-packed with an aluminum honeycomb panel, was placed at the top of each converter. The alignment unit represents the standard surface of the detector system, and each angle of a track recorded in the converter is calibrated by exploiting the high-momentum tracks that penetrate both the alignment unit and the converter. 
\par
The multi-stage shifter system was placed at the bottom of the converter and  used as time stamper.
Two or three emulsion films (38.8 $\times$ 25 cm$^2$ in size) were mounted on each movable stage.
Three stages were driven by stepping motors. 
During the observation period, each stage slid periodically with a different frequency, 
similarly to an analog clock, and created unique combinations of the stage position for any time period.
\par
The energy measurement for gamma rays above $\sim$ 1 GeV is performed  with a calorimeter.
It has a sandwich structure, which consists of 16 emulsion films and 15 1-mm-thick stainless-steel plates. 
The total thickness is 19.3 mm, which corresponds to 0.90$X_0$.
\par
The balloon also carries three daytime star cameras used as attitude monitors, as shown in Fig. \ref{detector}(c), a balloon-style pressure vessel, and several sensors (for a global positioning system, temperature, pressure, etc.) as its payload. 
 
\section{Emulsion film data acquisition and performance}\label{data-taking}
\begin{figure}
\begin{minipage}{0.4\hsize}
\begin{center}
\includegraphics[bb = 70 50 970 660, width=7cm,clip]{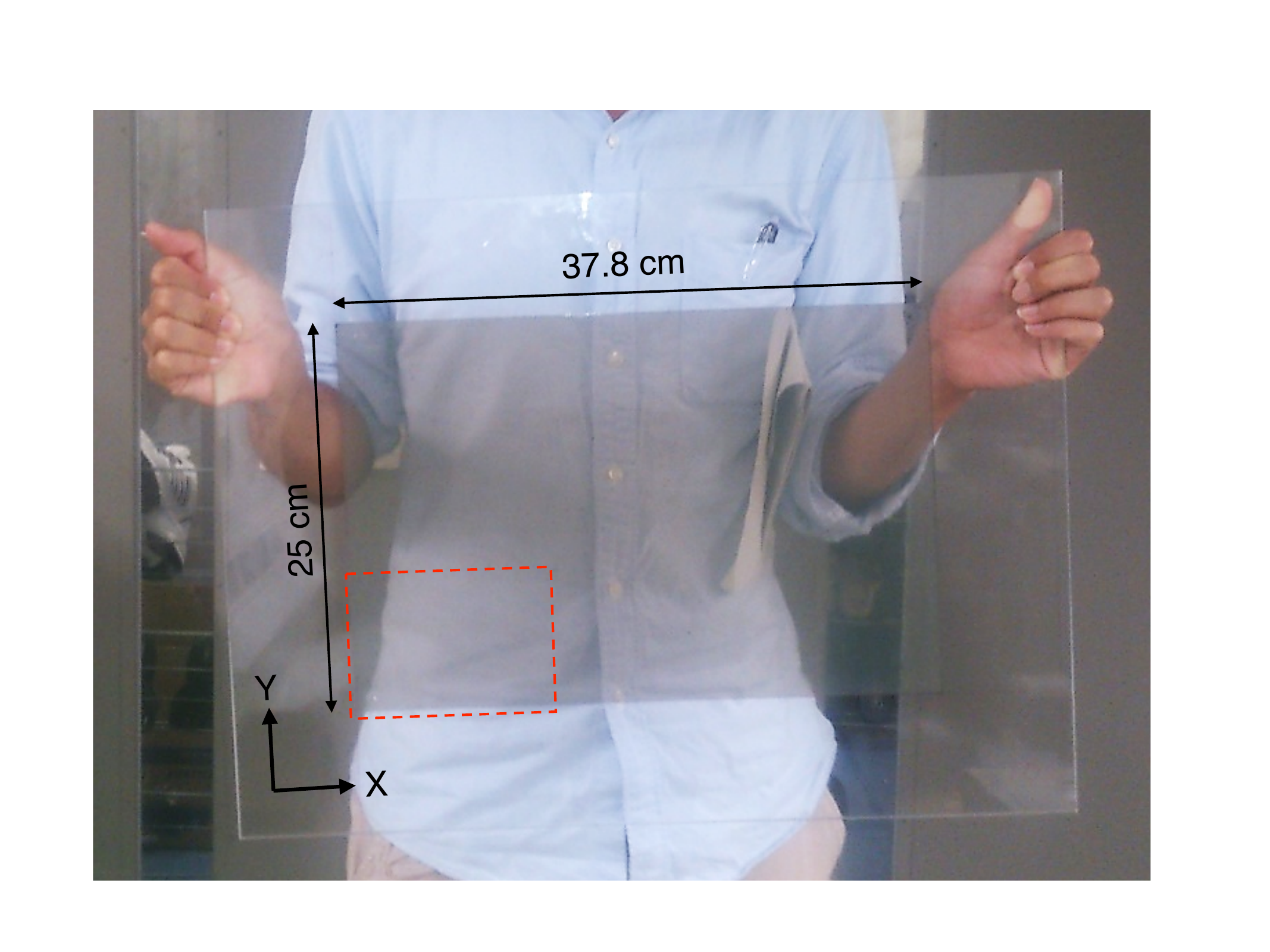}
\end{center}
\end{minipage}
\begin{minipage}{0.5\hsize}
\begin{center}
\includegraphics[width=9cm,clip]{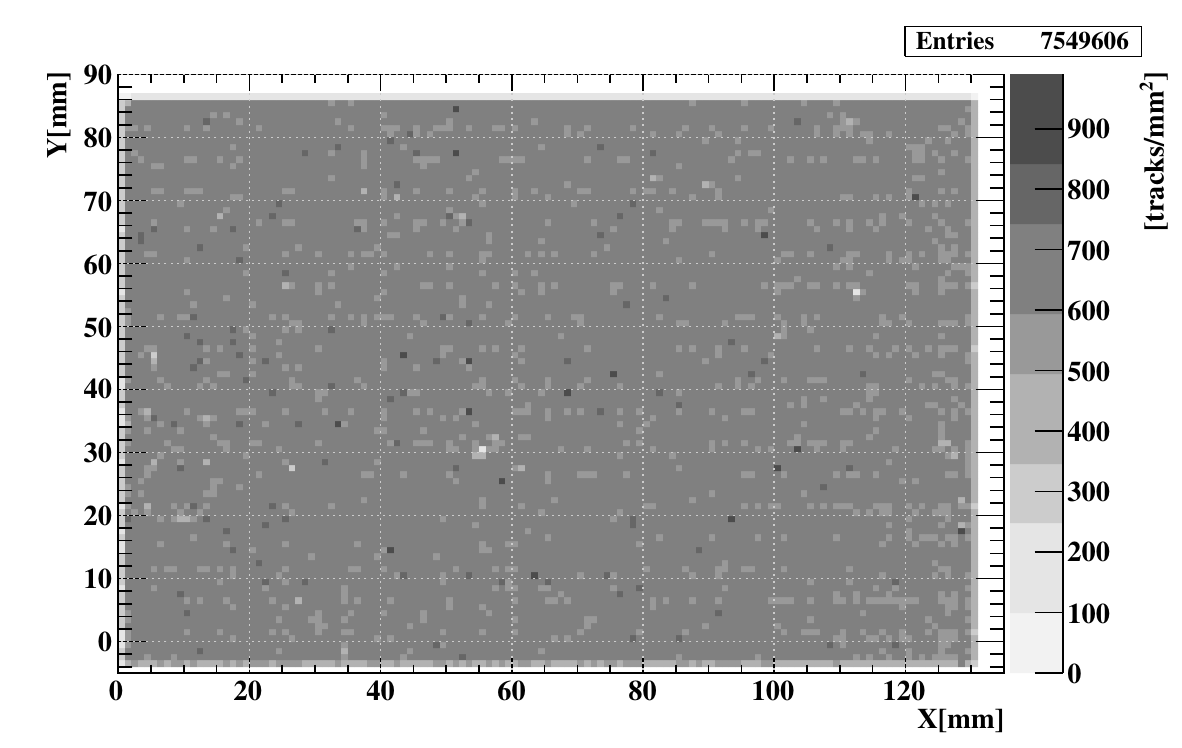}
\end{center}
\end{minipage}
\caption{
Film employed in the GRAINE 2015 experiment (left). The dotted square indicates the scan-unit area (13 $\times$ 9 cm$^2$). Position distribution of M-Tracks (right), which were scanned in a single emulsion layer of a film (unit \#3, film \#11, area \#5, TOP surface).
}
\label{scandata}       
 \end{figure}

\begin{figure}[]
\centering
\includegraphics[bb = 0 0 1100 700, width=12cm,clip]{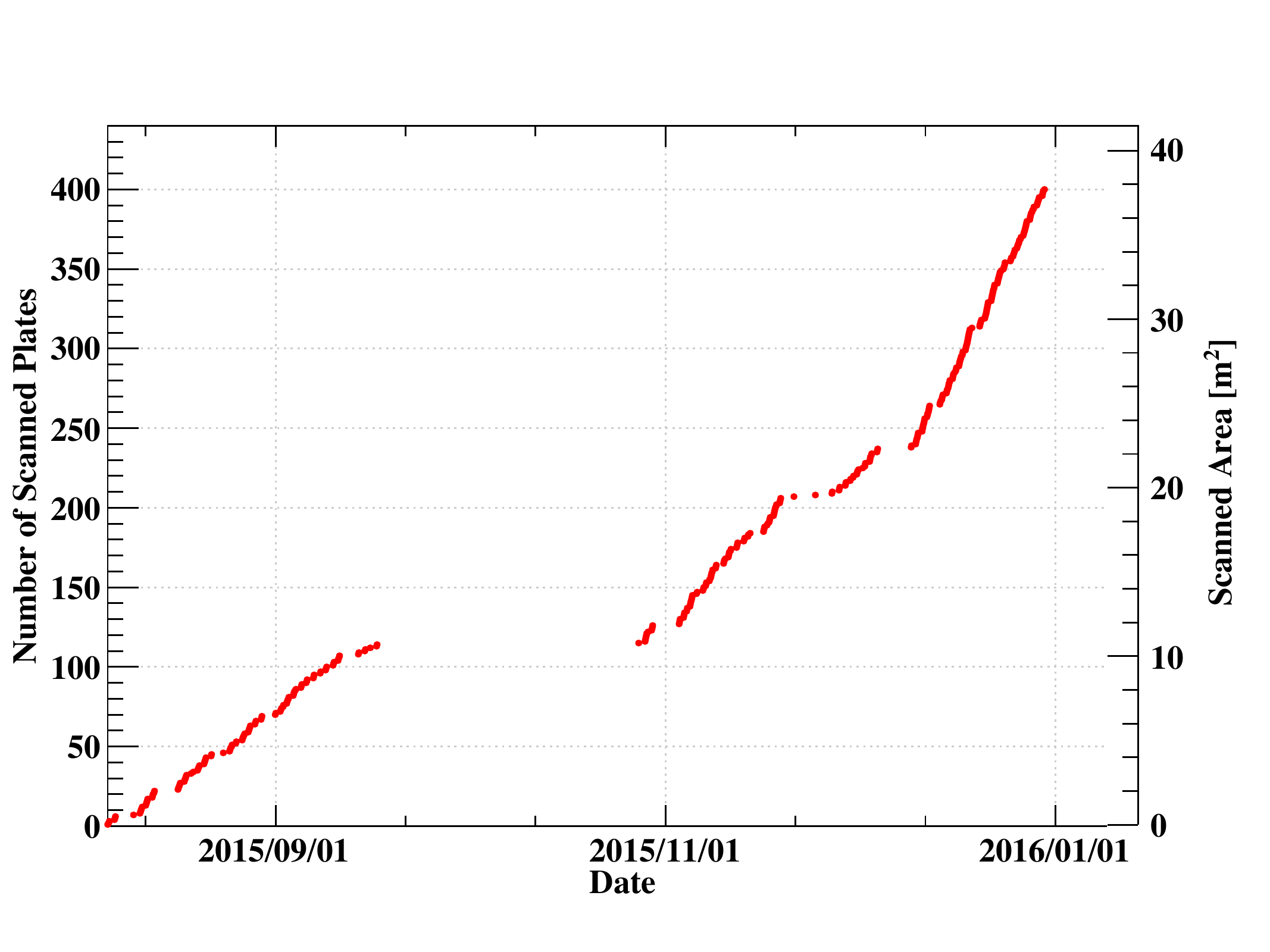}
\caption{
Number of scanned plates (left vertical axis) and scanned area (right vertical axis) of the converter films as a function of the time.
}
\label{scan}
\end{figure}

After the development and several treatments for the scanning (removal of surface silver and swelling), the latest scanning system, the Hyper Track Selector (HTS), which was developed in Nagoya University \cite{HTS}, was used for emulsion film data acquisition.
The angular range of the scanning process was set as
$|\tan\theta_{X}|<1.4$  and $|\tan\theta_{Y}|<1.4$, assuming that the angle of a vertical track on a film is defined as
$(\tan\theta_{X}, \tan\theta_{Y})=(0,0)$. 
The left photograph in Fig. \ref{scandata} shows the emulsion film that was attached to the acrylic plate to be set on the stage of the HTS.
Because the stage is movable by 13 cm and 10 cm in the x and y directions, respectively,  
we scanned a film with $3\times3$ divisions (each scan-unit area was set as 13 $\times$ 9 cm$^2$) to cover the entire area of the film.
The plot on the right-hand side of Fig. \ref{scandata} shows the density map of M-tracks obtained from a scan unit. As indicated in Fig. \ref{detector}(a), M-tracks are segments of a given track scanned on a single side of the emulsion layer of a film.
Figure \ref{scan} shows the number of scanned plates and the scanned area of the converter films as a function of the time.
The data acquisition for the GRAINE 2015 films was the first practical operation of the HTS.
All the films of the converter (37.8 m$^{2}$ in total) and the time stamper (3.5 m$^{2}$) were scanned in approximately three months, except in the periods dedicated to the update of the HTS.

In the GRAINE 2011, the OPERA films \cite{taku} and the data acquisition system developed for the OPERA experiment were employed. Due to the insufficient sensitivities of the emulsion films, the thresholds adopted in the find-track process were lowered in order to increase the track-finding efficiencies. However, this resulted into a poor S/N ratio as far as the M-tracks were concerned. Such a low S/N ratio was mitigated in the data analysis by considering the coincidence of M-tracks on both sides of the film (defined as B-tracks) and the coincidence of B-tracks between adjacent films (defined as L-tracks). We addressed this issue in the GRAINE 2015 experiment by introducing highly sensitive emulsion films; in this way, we did not only achieve a good track-finding efficiency, but also a reasonable S/N ratio, even for M-tracks. 
A summary of track-scan results, including number of tracks and efficiencies, is reported in Table \ref{film_table} for both GRAINE 2015 and GRAINE 2011. 
\par
\begin{figure}[]
\centering
\includegraphics[width=12cm, clip]{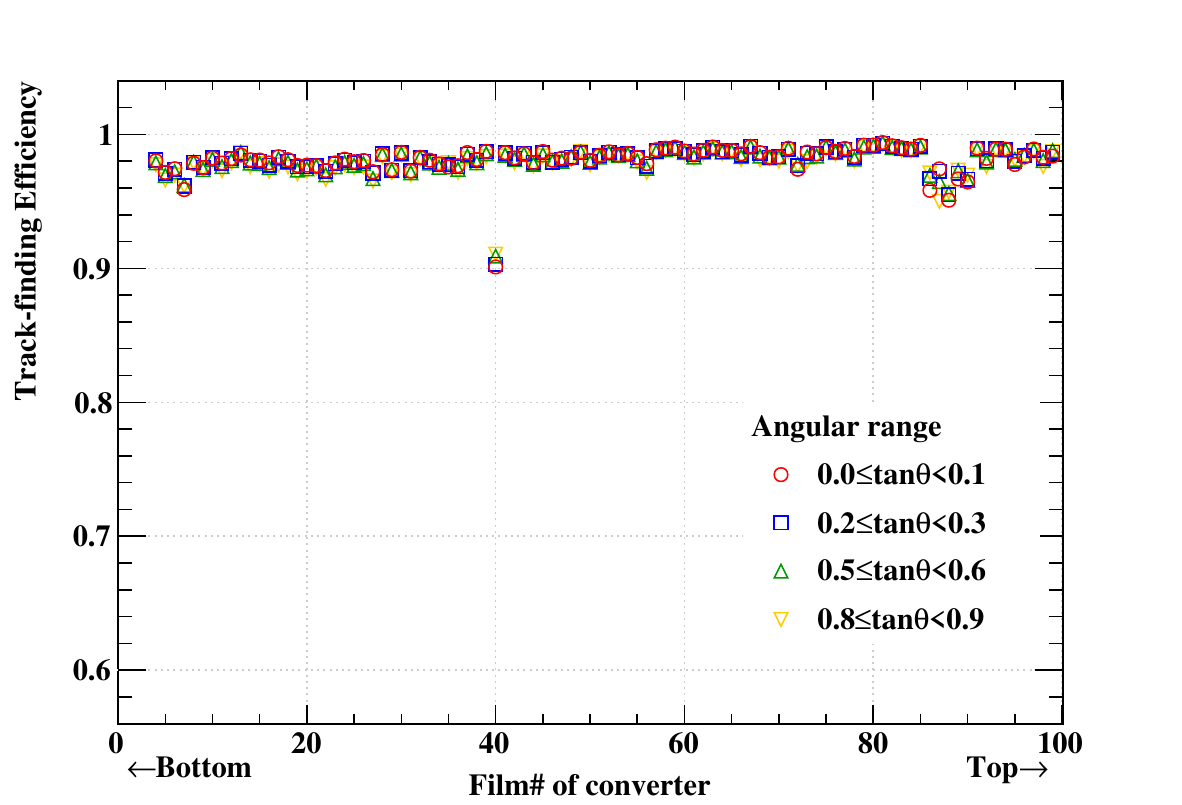}
\caption{
Track-finding efficiency as a function of the film number of the converter (unit \#3). }
\label{track-finding-efficiency}
\end{figure}
\begin{figure}[]
\centering
\includegraphics[width=12cm, clip]{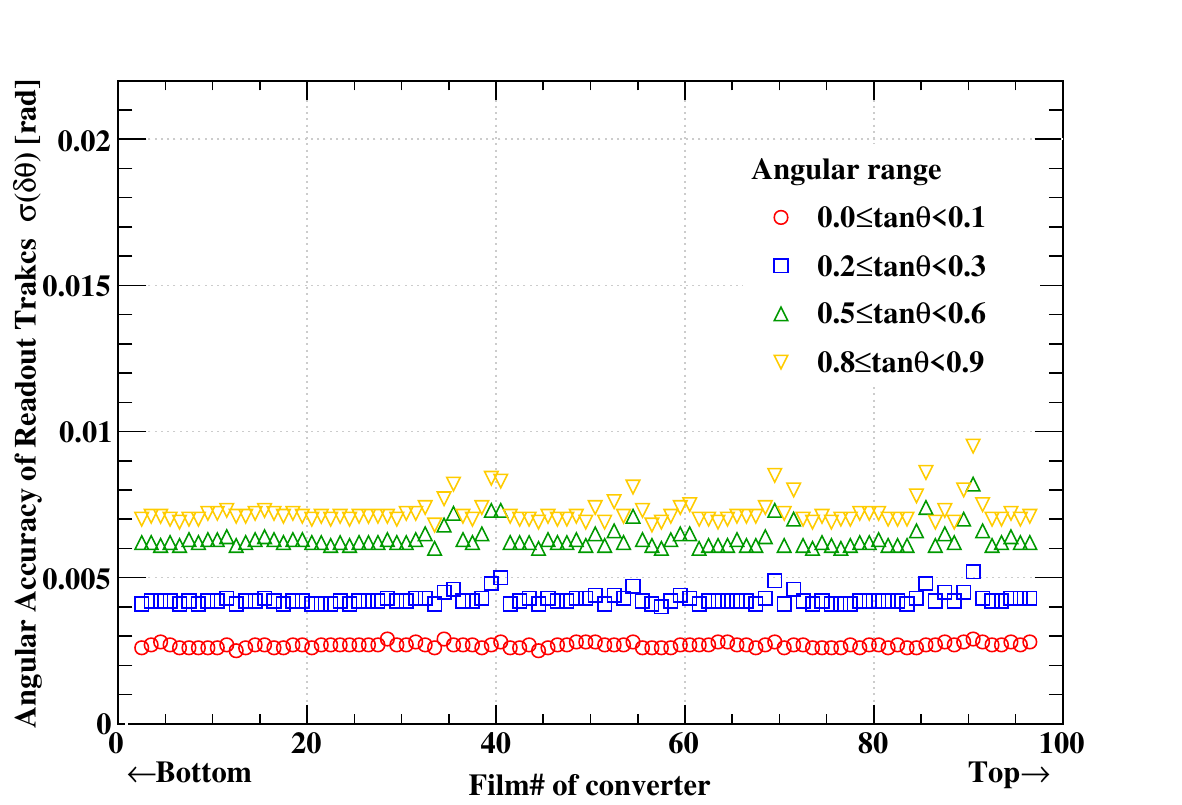}
\caption{Angular accuracy of the B-tracks evaluated in each film (unit \#3).}
\label{accuracy}
\end{figure}

The track-finding efficiency and the angular accuracy of M-tracks and B-tracks obtained by the HTS and GRAINE 2015 films have been previously studied \cite{ozaki, HTS}.
Figures \ref{track-finding-efficiency} and \ref{accuracy} show the track-finding efficiency and the angular accuracy of B-tracks, respectively; these results have been obtained by performing a film-by-film evaluation in one of the converter units.
The track-finding efficiency is defined as the ratio between the number of B-tracks found in the evaluated film data and the number of predicted tracks penetrating the two films upstream and downstream adjacent to the evaluated one.
The angular accuracy of B-tracks, indicated as $\sigma(\delta\theta)$, is defined as the sigma value of the distribution of angular differences between pairs of B-tracks connected to the adjoining film.
The track-finding inefficiency for B-tracks decreased by more than 75\% compared with that of the GRAINE 2011 experiment. 
However, a few films with irregularities were observed.
These were mainly caused by human errors in the swelling, which is the process of soaking emulsion layers into the glycerin in order to recover their thickness shrunk after the development. Overswollen films  showed slightly lower track-finding efficiency and accuracy. 
However, the basic performances of the majority of the films of the converter are stable in each angular range and in each film, thus allowing a systematic analysis.

\begin{table}[t]
\caption{
Summary of the characteristics of emulsion films employed in GRAINE 2011/2015.
}
\label{film_table}
\centering
\begin{tabular}{l|rr}
 & GRAINE 2011 \cite{GRAINE2011} & GRAINE 2015\\ 
\hline\hline
Type of emulsion film		& OPERA film	& Nagoya-gel film\\
Read-out system &	S-UTS	&	HTS \\
Angular range ($|\tan\theta_{X/Y}|)$ &	$<1.0$	&	$<1.4$\\
\hline
Scanned Area (m$^2$)&	&	\\
Converter 　			&	1.25& 37.8	\\
Timestamper　			&	0.1&	3.5\\

\hline
Number of tracks (/125 cm$^{2}$)& & \\
M-tracks  &$2\times10^{8}$&  $8\times10^6$\\
B-tracks  &$1\times10^{7}$&  $5\times10^6$  \\
L-tracks  &$1\times10^{6}$&   $4\times 10^6$\\
\hline
Track-finding efficiency for B-tracks & 80\%	&	$>$95 \%\\
\end{tabular}
\end{table}

\section{Analysis for gamma-ray event selection} \label{Analysis-for-selection}
The gamma-ray event selection from a balloon-borne converter was performed in the data analysis of  the GRAINE 2011 experiment \cite{GRAINE2011}. 
However, because of the poor efficiency and the low S/N ratio of the scan data, the analysis load was heavy. 
In the GRAINE 2011 analysis, we attempted to search for electron-pair events that occurred at the seventh film from the bottom of the converter (the size of the film was 12.5 $\times$10.0 cm$^2$). We selected 153 events of this kind.
However, in order to distinguish the signal (153 events) from the background ($10^3$ events), in the final process of the selection it was necessary to analyze by eye the event topologies on the display of a three-dimensional (3D) event viewer. In the GRAINE 2015 analysis, we selected gamma-ray events without recurring to any human-eye check, but instead using a high-quality data selection method, as described in the following section. 

\subsection{Automatic selection for the conversion points and track follow down (TFD)}
\begin{figure} 
\centering
\includegraphics[bb = 0 100 1020 700, width=15.5cm,clip]{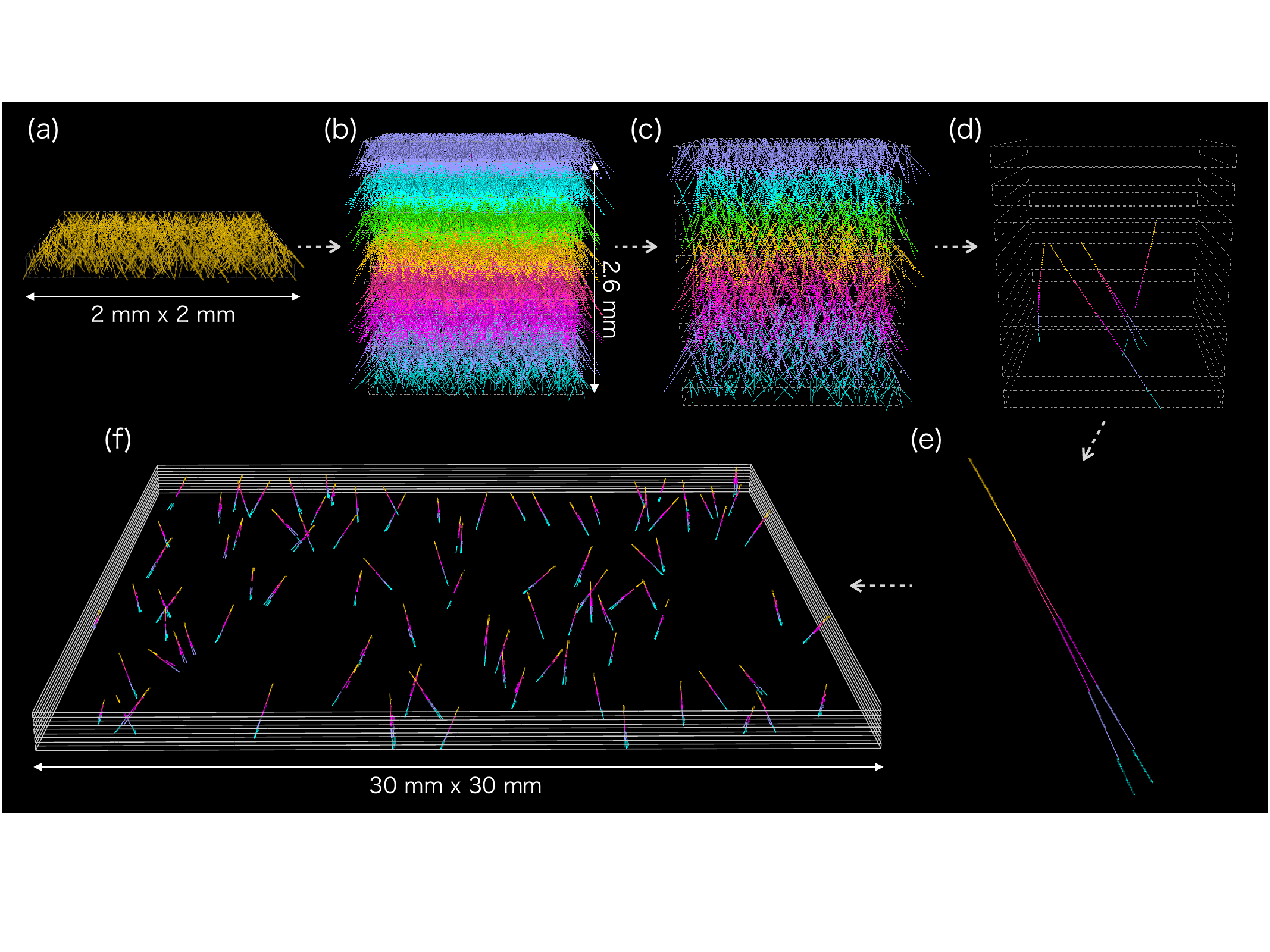}
\caption{
Flow of the automatic selection of conversion points using the real data of the GRAINE 2015 experiment: (a) 2 $\times$  2 mm$^2$ area in a GRAINE 2015 film; (b) volume of 8 films; (c,d) datasets after that penetrating tracks are eliminated and required to start at the target film, respectively; (e) typical gamma-ray event detected by the selection process; (f) wide view (30 $\times$ 30 mm$^2$) resulting from gamma-ray detection.
}
\label{selection}       
\end{figure}

\begin{figure}[]
\centering
\includegraphics[width=13cm, clip]{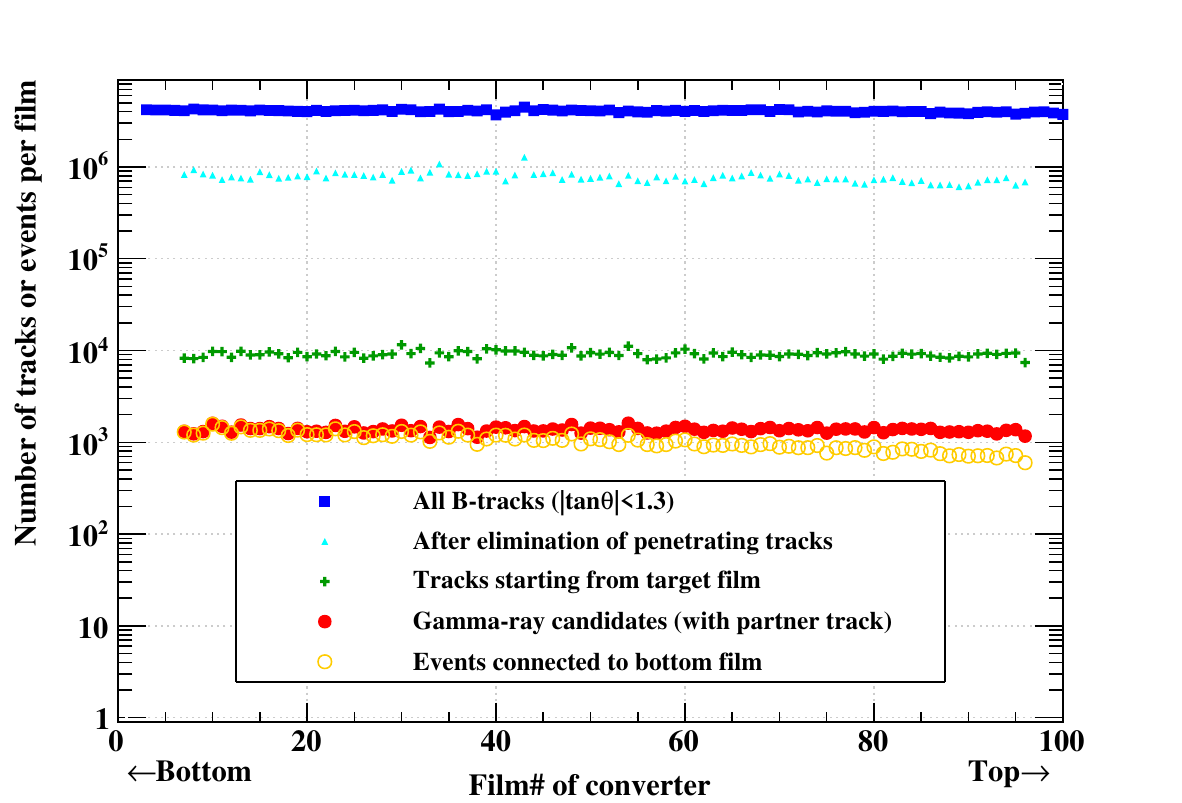}
\caption{Results of gamma-ray event ($\gamma + (\gamma) \rightarrow$ e$^+ + $e$^-$) selection using process-unit data from the converter (13 cm $\times$ 9 cm $\times$ 100 films)}
\label{reduction}       
\end{figure}

Figure \ref{selection} shows the flow of the automatic selection of a conversion point by using the real data of the GRAINE 2015 experiment. 
In more details, we searched for conversion points that start from a ``target film'' and by following the steps described below.
Then, we implemented the same process on almost the entire volume of the converter, while changing the target film from the downstream region to the upstream region. 
The steps were: (a) 400 B-tracks were scanned by the HTS from a 1-mm$^2$-wide area of a film in the angular region defined by $|\tan\theta_{X}| < 1.3$ and $|\tan\theta_{Y}| < 1.3$;
(b) a dataset consisting of 8 films (a target film, three films upstream from the target film, and four films downstream from the target film) was used to search for conversion points;
(c) the tracks that penetrated the volume of the 8 films were eliminated.
The ``NETSCAN'' method was adopted to reconstruct chain tracks, where a chain is defined as a track consisting of a series of connections of B-tracks \cite{NETSCAN, hamada}.
Chain tracks with more than 6 B-tracks in whole volume of the 8 films were removed from the dataset as penetrating tracks.
After this process, the number of tracks decreased to approximately 20\% of the original number of B-tracks; 
(d) chain tracks starting on the target film and reconstructed in the 4 downstream films were selected. After the requirement of more than 4 B-tracks to reconstruct chain tracks, the number of tracks decreased to approximately 1\%;
(e) the paired topology, wherein another independent track runs abreast nearby the track, was requested. Position and angular differences between two tracks were set as $\Delta r<50$ $\mu$m and $\Delta\tan\theta<0.15$, respectively, where $\Delta r=(\Delta  x^2+\Delta y^2)^{1/2}$ and $\Delta \tan\theta = (\Delta\tan^2\theta_{X}+\Delta\tan^2\theta_{Y})^{1/2}$.
The number of remaining tracks was approximately $10^3$ in the target film.
Figure \ref{reduction} shows the result of the gamma-ray event selection using the process-unit data of the converter (13 cm $\times$ 9 cm $\times$ 100 films).
\par

\begin{figure} 
\centering
\includegraphics[bb = 0 0 870 750, width=10cm,clip]{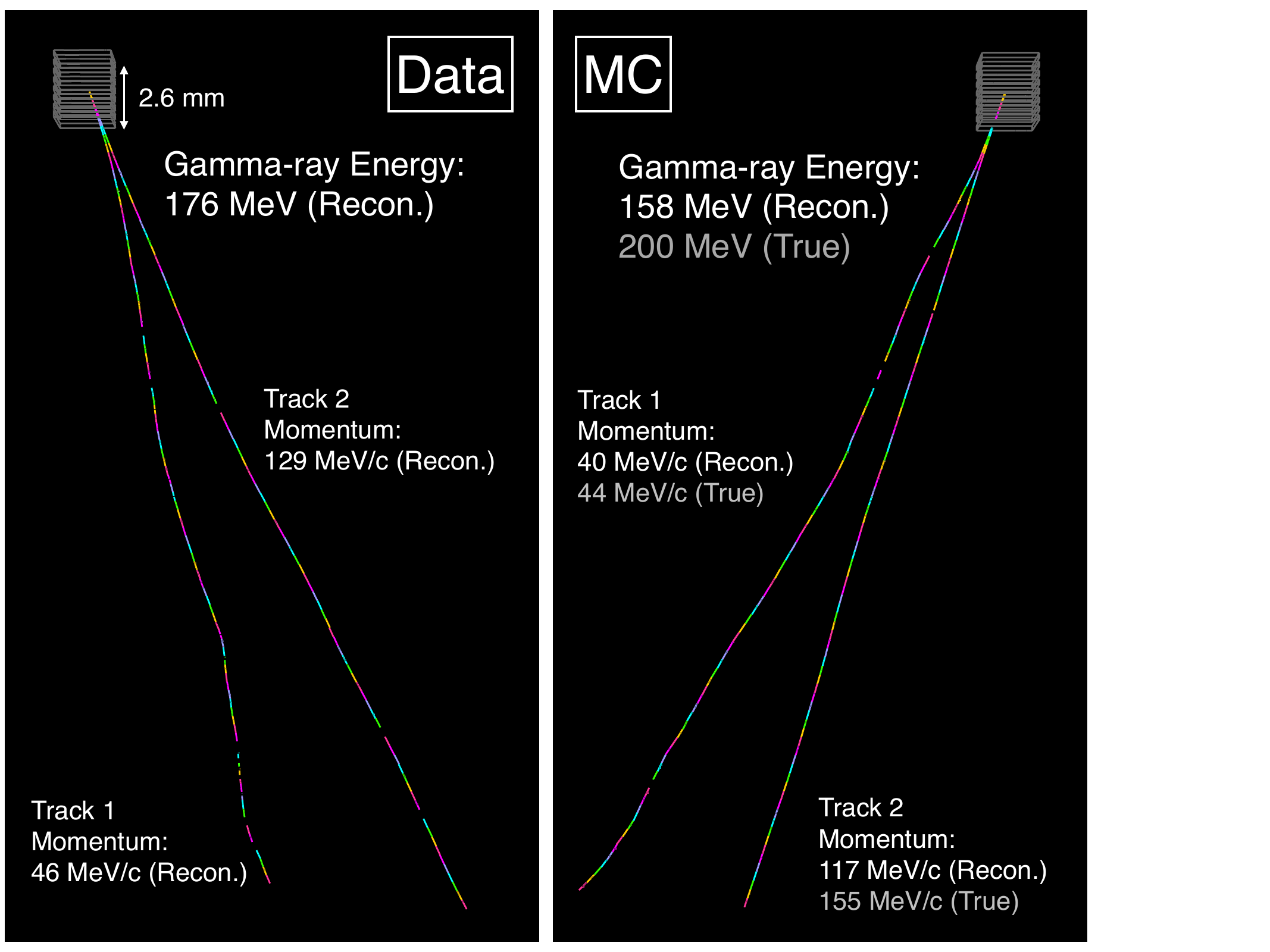}
\caption{Typical events as results from the TFD. The figure on the left shows the flight-data event that occurred at film \#95. The figure on the right shows the event generated by MC simulation with similar gamma-ray energy.}
\label{FDview}       
\end{figure}

These electron and positron tracks were connected downward from the target film, according to the method called track follow down (TFD).
The events that reach the bottom film of the converter are the ones suitable for a time-stamp analysis.
In case of an event that starts in the most upstream part of the converter, more than 90 connections across the films are needed. 
The TFD method was performed by looking for the unique B-track that matches the angle and the position, and by repeating connections from one film to the other. When multiple B-tracks to connect are found, the best connection is selected via the maximum likelihood estimation method \cite{Fukuda}; when no candidate is found in three consecutive films, the TFD is stopped. 
Figure \ref{FDview} shows the TFD results for an event that starts from film \#95.
After the TFD process, 93700 events were selected as effective gamma-ray events in a process unit.
The efficiency of the event selection and the background contamination are discussed in section \ref{purity_eff}.

\subsection{Energy reconstruction in converter}
The gamma-ray energy is reconstructed by measuring the momenta of electron and positron tracks \cite{GRAINE2011}. The momentum can be reconstructed from multiple Coulomb scattering (MCS) processes in the converter and/or the calorimeter. 
The difference in momenta between the two tracks in Fig. \ref{FDview} is indicated by scattering angles.
\par
\begin{figure}[]
\centering
\includegraphics[width=11cm,clip]{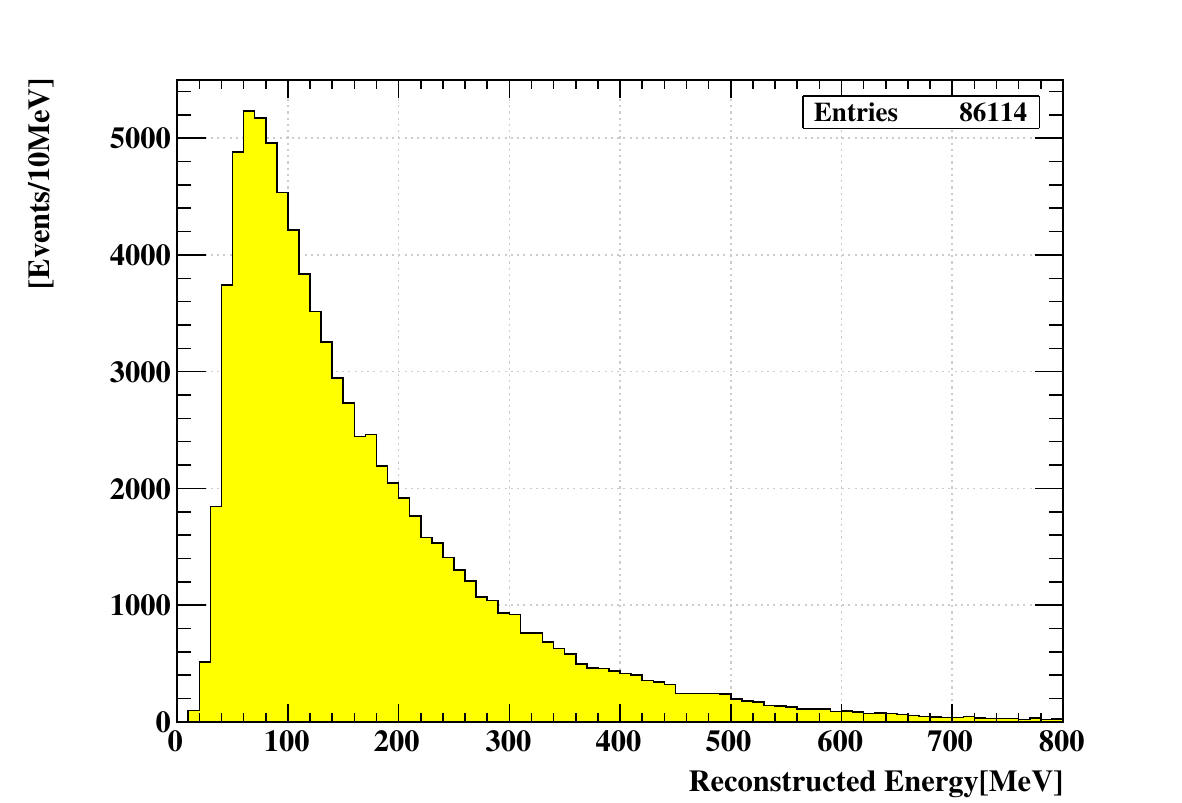}
\caption{Distribution of the gamma-ray energy measured in the converter. The events are selected in a process unit (13 cm $\times$ 9 cm $\times$ 100 films).  }
\label{energy}       
\end{figure}

 
Energy reconstructions were performed for all the 93700 events selected after the TFD. In this analysis, we used the angles of B-tracks near the conversion point given by the result of the TFD, and we applied the angular method to reconstruct the momenta \cite{Park}. 
In the angular method, the multiple angle differences are measured by sampling the segmented tracks (here, B-tracks) from a chain track, and the momentum is estimated from the root-mean-square of the angle differences in a track. 
In the measurement of the angle difference, the path length between two B-tracks is called cell length.
We set the maximum cell length at 2 in this analysis (a unit of cell length is defined as the number of emulsion films between B-tracks). 
As a result, finite energies were reconstructed for 86114 events (the fraction of successes is 92\%).
Figure \ref{energy} shows the distribution of the reconstructed gamma-ray energy.
Lower limits, in the case of no significant MCS observed in either one or both electron/positron track(s), were set for 7586 events (8\%). 
\par
\begin{figure} 
\centering
\includegraphics[width=11cm,clip]{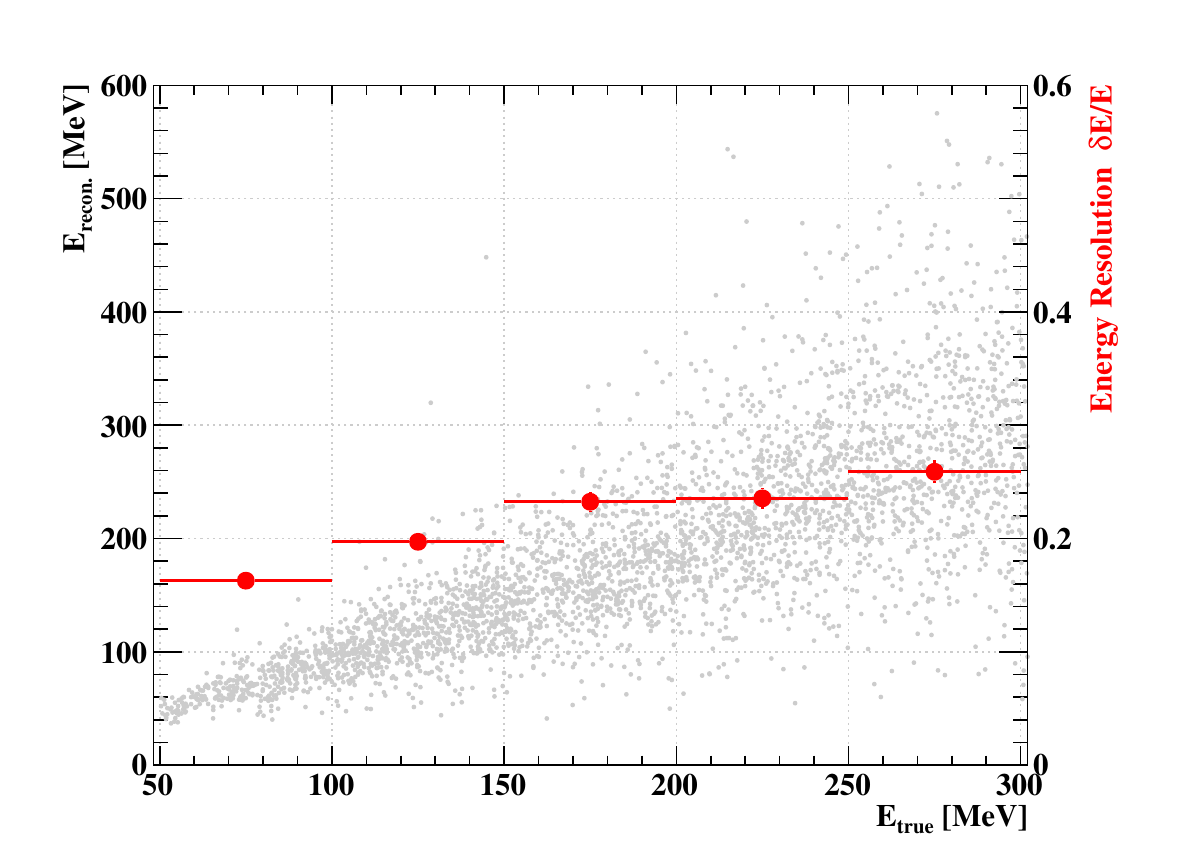}
\caption{Energy resolution at the converter under the GRAINE 2015 conditions (the maximum number of independent angle was set at 20 in each projection and the maximum cell length was set as 2), which was evaluated by MC simulation.}
\label{MomSim}       
\end{figure}

Figure  \ref{MomSim} shows the energy resolution estimated by Monte Carlo (MC) simulation in the condition of the GRAINE 2015 analysis (the maximum number of sampled angles was set at 20 in each projection).
MC data were generated by Geant4.10.01 \cite{Geant} (the MC data presented in the following section are generated by the same version of Geant4).
The gray scatter plot reports the reconstructed energy ($E_{recon.}$) and the true energy ($E_{true}$) on the left vertical axis and on the horizontal axis, respectively. 
The red points indicate the relative resolution $\delta E/E$ (right vertical axis), where $\delta E$ is the sigma value, i.e., the result of the Gaussian fit of the residual distribution ($E_{recon.} - E_{true}$) in each energy range.
The energy resolution is mainly limited by the statistics of the sampling number of scattered angles.
For those electrons and positrons that have traveled over long distances, the effect of bremsstrahlung and energy loss by ionization cannot be ignored.
However, the average contribution of such processes can be calculated from the amount of material that electrons and positrons have passed starting from the conversion point. 
It is expected that, by considering these effects, the resolution of the initial momentum will be improved by sampling more B-track angles further downstream. 
Moreover, the fraction of the events with only lower energy limits can be decreased by considering larger cell length in the analysis, and/or adopting the coordinate method, which uses the position displacement between films and is suitable for the measurement of high-momentum particles \cite{Park}.
 Another possibility would be performing the analysis in the calorimeter, which presents thicker material (radiation length) between films.
\par

\subsection{Estimation of background contamination and efficiency for event selection}\label{purity_eff}
We estimated the background contamination for the gamma-ray event selection.
Background events are classified into chance-coincidence background (CC-BG) and hadronic interaction background (HI-BG).
CC-BG is due to two independent tracks that start from the middle and satisfy the topological criteria by chance.
To evaluate the number of CC-BG events, $10^4$ tracks starting on a film were picked up, 
and the search for a partner track was performed with the same criteria adopted for the topology selection, after having randomly changed the position of the starting track. As a result, 30 fake events were found, corresponding to 0.3\% of the seed tracks.
Typically, 1400 events with a partner track were selected from 9000 starting tracks in the target film (Fig. \ref{reduction}), and the number of CC-BG events was estimated to be 9000 × 0.003 = 27 events, corresponding to 1.9\% of the track pairs. 
When high-energy cosmic rays hadronically interact with the converter, multiple secondary particles are generated at the vertex. Two of them with similar angles sometimes satisfy the signal criteria; thus, we define such events as HI-BG.
For the HI-BG events, an additional search was performed on 500 pair-topology events to determine whether the selected tracks converge with secondary track(s) and a primary track or not.
As a result, 17 events were found, and the HI-BG fraction was estimated to be $\frac{17}{500}=3.4\%$.
Therefore, the total background contamination fraction is 5.3\%. 
\par
\begin{figure}[]
\centering
\includegraphics[width=11.0cm,clip]{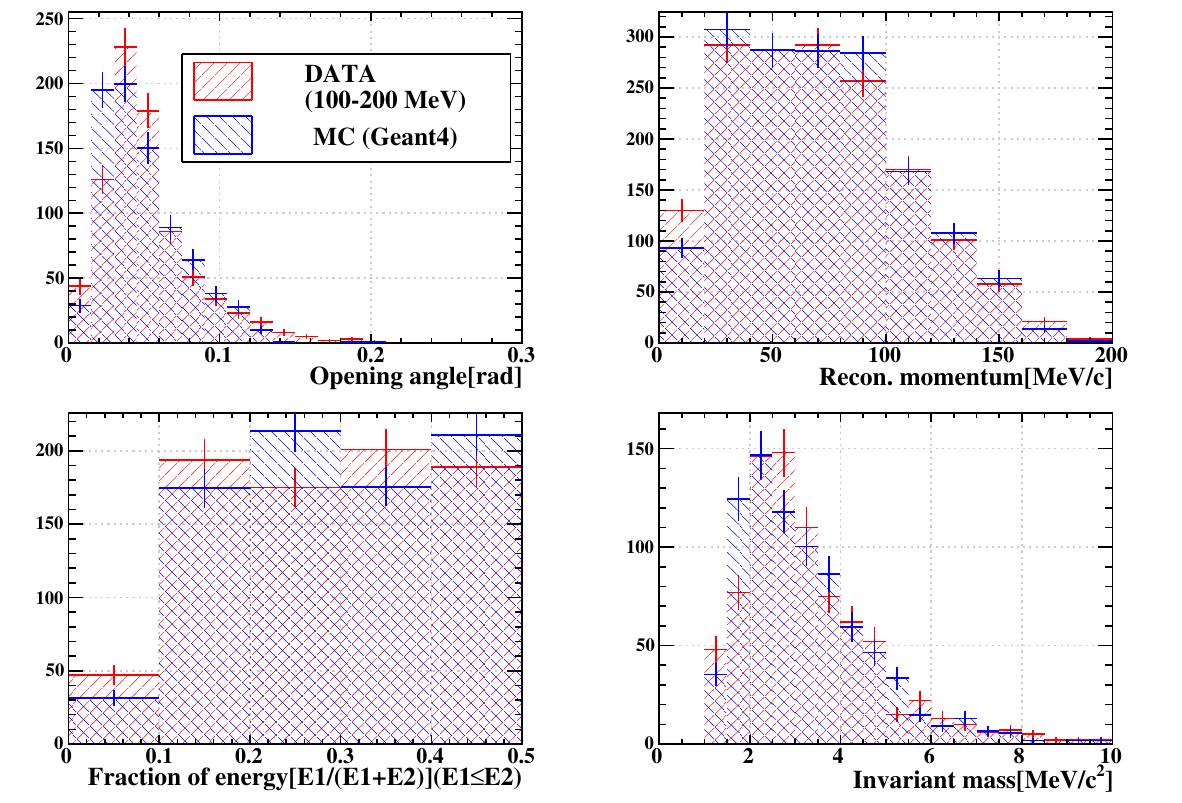}
\caption{Comparisons of the kinematical distributions of electron and positron pair tracks (100--200 MeV) between flight data and MC simulation.}
\label{DataMcPlot3}       
\end{figure}

We checked the response of the selection by MC simulation.
The detector was exposed to incident gamma rays, and then the response of HTS (track-finding efficiency, position accuracy, and angular accuracy) was applied to
the electron and positron tracks produced in the emulsion films.
A simple power-law spectrum was employed as the energy distribution of incident gamma rays.
The topological selection was performed similarly to the aforementioned analysis.
Figure  \ref{DataMcPlot3} shows the distributions of the opening angle between electron and positron tracks, the reconstructed momentum, the fraction of the energy for the lower-energy track with respect to the total energy, and the invariant mass.
 The MC data reproduced the trends of the flight data; however, flight data show a slight bias toward higher opening angle, and therefore higher invariant mass, due to the incomplete application HTS parameters for the clustering of close tracks.
\par

\begin{figure}[]
\centering
\includegraphics[width=11.0cm,clip]{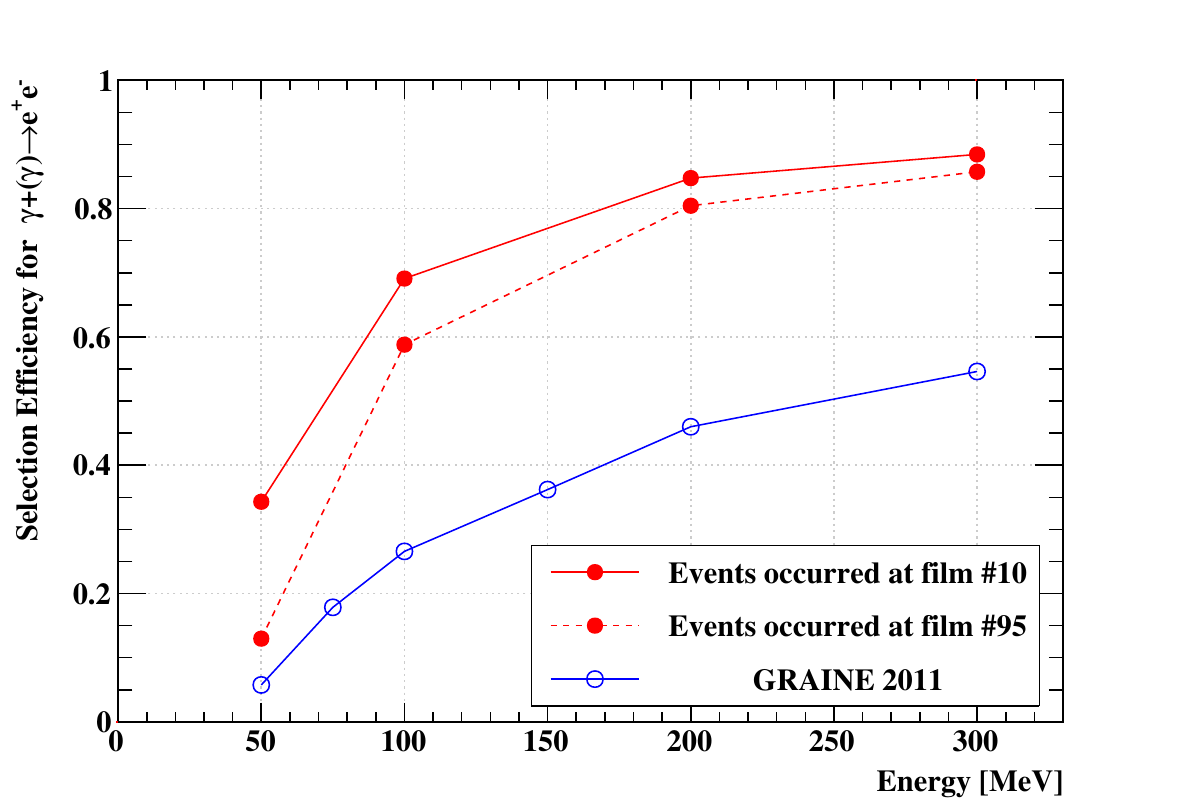}
\caption{Selection efficiency  for $\gamma + (\gamma) \rightarrow e^+e^-$ events in the converter, estimated by MC simulation.}
\label{DetectionEfficiency}       
\end{figure}

We define the selection efficiency  for $\gamma + (\gamma) \rightarrow e^+e^-$ events in the converter as the success of selecting conversion points and connecting one between the electron and positron, or both, to the bottom film after TFD.
Figure \ref{DetectionEfficiency} shows the energy dependence of the selection efficiency evaluated by MC simulation. As the TFD efficiency depends on the depth of the conversion point, the results for upstream (film \#95) and downstream (film \#10) events are separately indicated in Fig. \ref{DetectionEfficiency}.
 In the GRAINE 2015 analysis, the selection efficiency was estimated to be 65\% and 83\% (averaged values for the depth of the chamber) for 100- and 200-MeV gamma rays, respectively.
This shows the substantial improvements with respect to the GRAINE 2011 analysis. 
\par
In the bottom-left distribution of Fig. \ref{DataMcPlot3}, a decreasing number of detected events is observed for an energy fraction less than 0.1. This is caused by the insufficient track-connection efficiency for $\sim10$ MeV electrons in the current algorithm. To reduce the background fraction and increase the efficiency, we will polish and further develop both the track-connection algorithm, along with the selection criteria using kinematical parameters, and the method to detect tracks from hadronic interactions in the balloon-borne emulsion chambers.

\subsection{Measurement of atmospheric gamma rays}
The gamma-ray events selected in the converter were connected to the time stamper and their corresponding incident times were evaluated (the time-stamp analysis is described in detail in \cite{Mizutani}).
 The main backgrounds for the observation of cosmic gamma rays consist of external components, i.e., atmospheric gamma rays, and internal components, i.e., secondary gamma rays induced in the detector by protons, electrons, etc. (cosmic-ray positrons can also produce gamma rays via annihilation, but the amount is incomparably small).
 The internal component can be rejected by identifying the hadronic interaction vertex or the electron running abreast with the same incident time as the gamma ray. 
The demonstration analyses for both the GRAINE 2011 and 2015 experiments are described elsewhere \cite{GRAINE2011, Kawahara}.
\par
\begin{figure}[]
\centering
\includegraphics[width=12cm, clip]{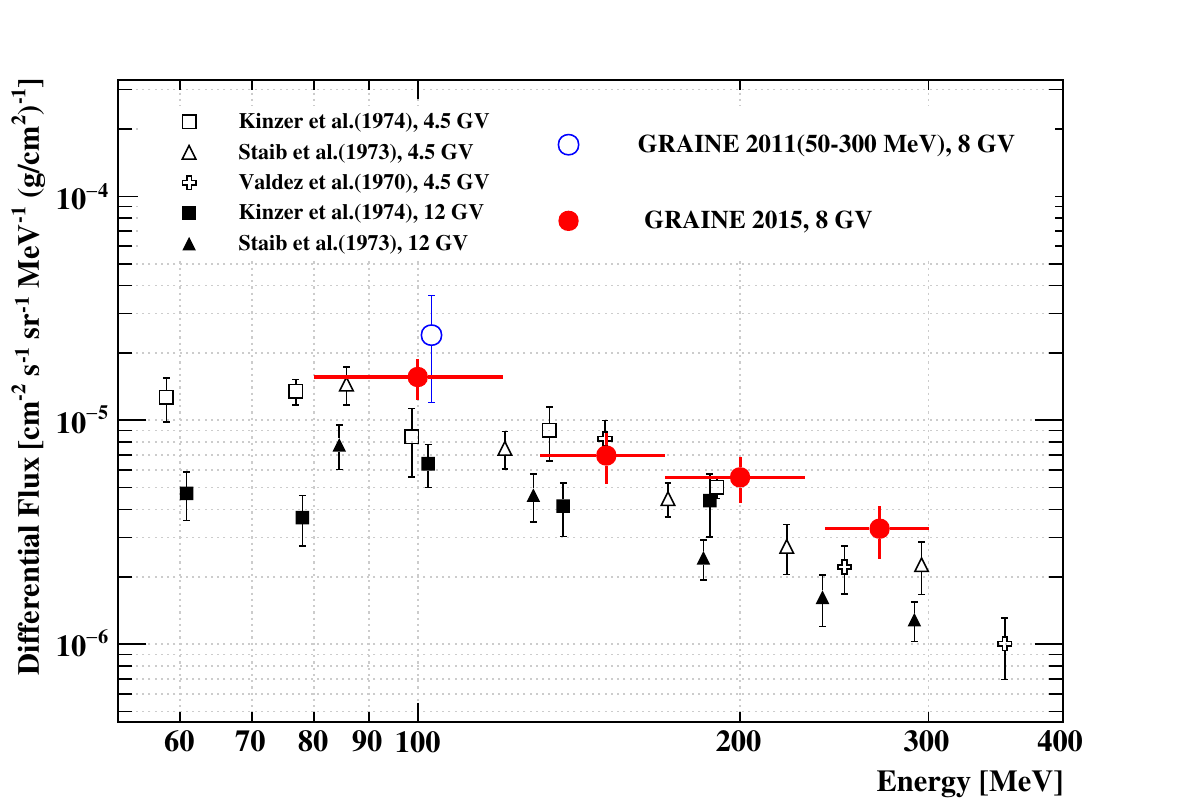}
\caption{Results of the measurements of the atmospheric gamma-ray flux.}
\label{flux}       
\end{figure}

Here, the atmospheric gamma-ray flux was measured in the altitude range of 36.0--37.4 km.
Gamma-ray events were accumulated from the upstream events  in 4 process units, which started from the 95th or upper films in the converter, to avoid the contamination from internal components.
Figure \ref{flux} shows the results of the measurement of the atmospheric gamma-ray flux in the energy range of 80--300 MeV. 
The vertical and horizontal error bars indicate statistical errors and energy ranges for each flux, respectively. Each flux is calculated by taking into consideration the migration of some detected events due to the energy resolution, and assuming that the energy spectrum obeys a simple power law around each energy region.
We observed a tendency similar to that of previous measurements. 
\par

\section{Demonstration of gamma-ray imaging during the balloon flight}\label{demonstration-of-imaging}

\subsection{External gamma-ray source for imaging demonstration}
\begin{figure}[]
\centering
\includegraphics[bb = 200 0 1700 1030, width=12.5 cm, clip]{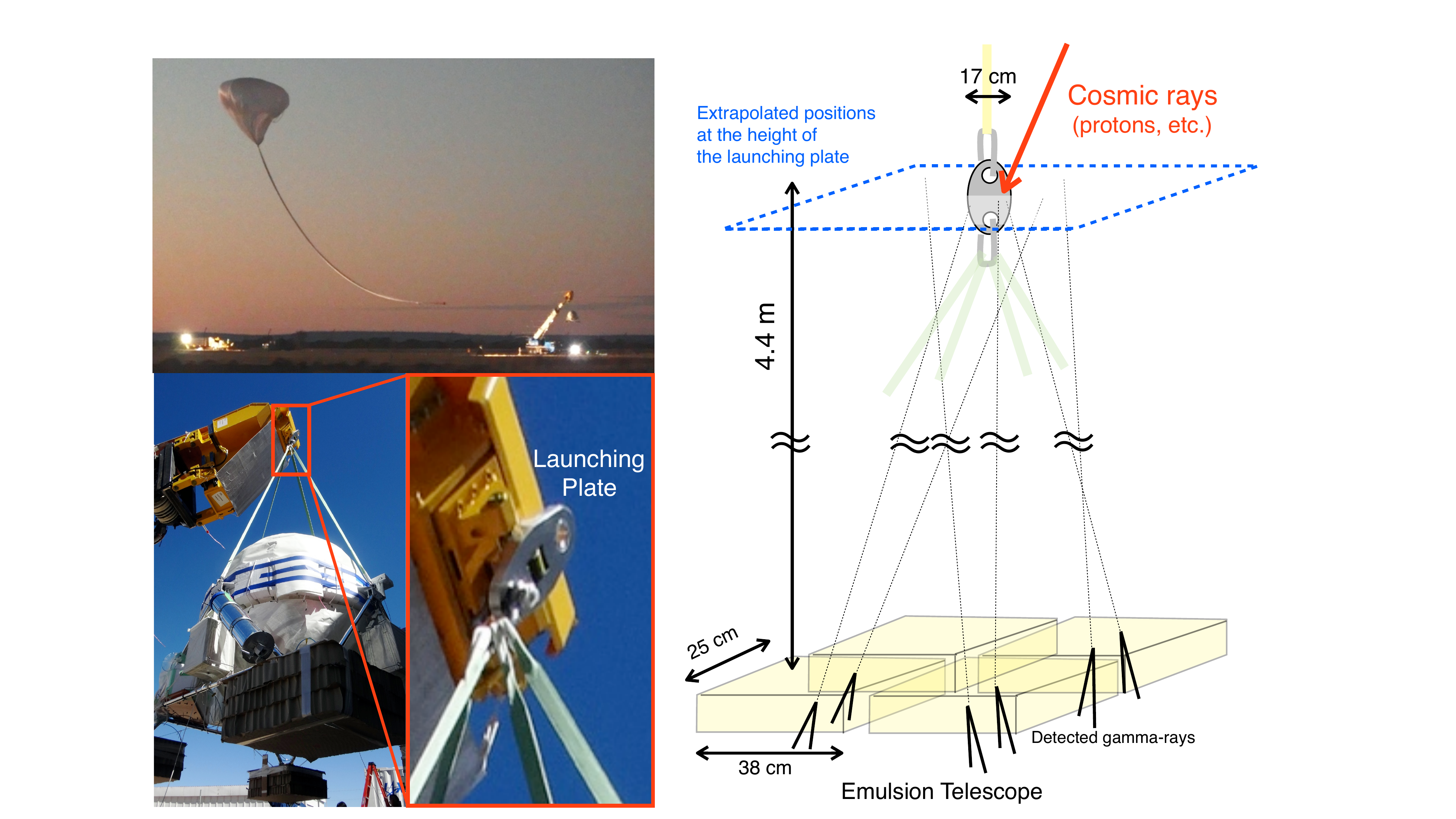}
\caption{
Photographs of the balloon, the crane truck, the gondola, and the launching plate (left). Schematic view of the positional relationship between the emulsion telescope and the launching plate (right). The square drawn with blue broken lines indicates a plane at the height of the launching plate.
}
\label{plate}       
\end{figure}

Figure \ref{plate} shows the photograph of the gondola suspended by a crane truck and the photograph of an aluminum plate, the launching plate that connected the payload to the large balloon.
The crane held the launching plate and released it at the time of the launch.
After the launch, the plate was situated 4.4 m above the emulsion telescope.
The plate was exposed to cosmic rays, mainly protons, at the observation altitudes.
When cosmic rays interact with the materials, secondary particles, including gamma rays, are created.
Thus, the launching plate became an external gamma-ray source for the calibration of the emulsion telescope during the flight.

\subsection{Result of gamma-ray imaging with the emulsion telescope}
\begin{figure}[]
\centering
\includegraphics[width=15.0 cm,clip]{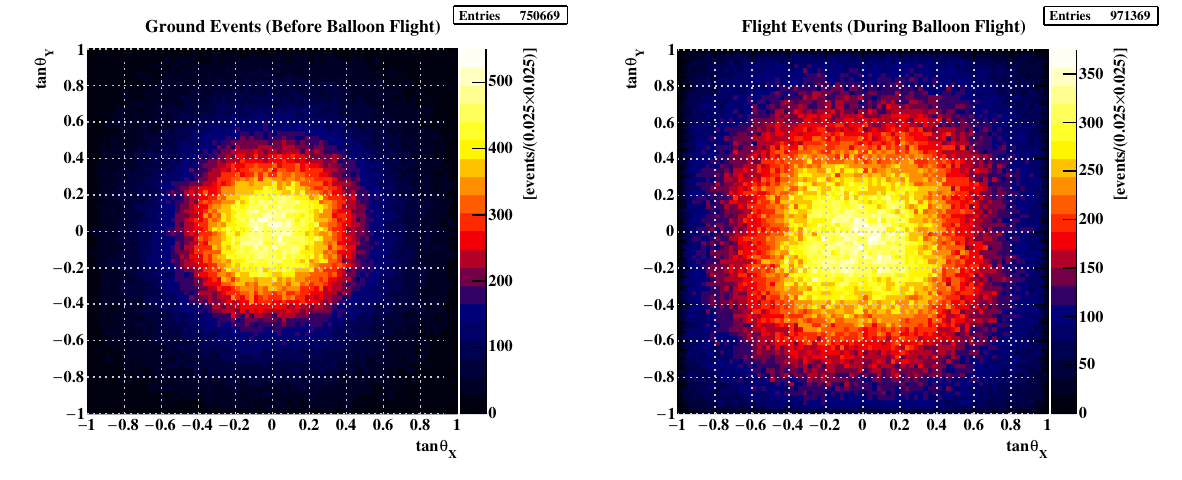}
\caption{
Angle distributions of gamma rays accumulated in the converter. The origin of angles in the detector coordinate is nearly equal to the direction of the zenith. The left figures show the ground events, which were incident before the balloon flight; the right figures show the flight events, which were incident during the balloon flight. 
 }
\label{angle_dist}       
\end{figure}

In order to create the gamma-ray image, we processed the automatic event selection for 27 process units, corresponding to an aperture area of 0.28 m$^2$.
As a result, 2$\times10^{6}$ gamma-ray events were accumulated. 
Figure \ref{angle_dist} shows the gamma-ray angle distributions in the detector coordinate.
The events were divided into ground and flight events based on the time stamp.
The origin of the angles is nearly parallel to the zenith direction.
The distribution of the ground events shows bias toward the zenith, which was due to the events arising from muons at the ground level via bremsstrahlung.
\par
The gamma-ray image of the launching plate is not expected to be focused on the angular distribution of selected gamma-ray events, because the aperture size of the emulsion telescope is not negligible with respect to the distance of the launching plate. 
Therefore, we mapped the extrapolated positions, which are the cross points between gamma-ray vectors and the plane at the height of the launching plate (see the right schematic view in Fig. \ref{plate}).
The extrapolated positions $X_{ext.}$ and $Y_{ext.}$ are determined by using the following equations:
\begin{equation}
X_{ext.} = \tan\theta_{X} \times (Z_{cnv.} + L) + X_{cnv.}
\label{extx}
\end{equation}
\begin{equation}
Y_{ext.} = \tan\theta_{Y} \times (Z_{cnv.} + L) + Y_{cnv.}
\label{exty}
\end{equation}
where the 3D position ($X_{cnv.}$, $Y_{cnv.}$, $Z_{cnv.}$) corresponds to a conversion point in the detector coordinates, $\tan\theta_X$ and $\tan\theta_Y$ are the projection angles of a gamma ray, and $L$ is the height of the launching plate from the surface of the detector.
\par
\begin{figure}[]
\centering
\includegraphics[width=15.0 cm,clip]{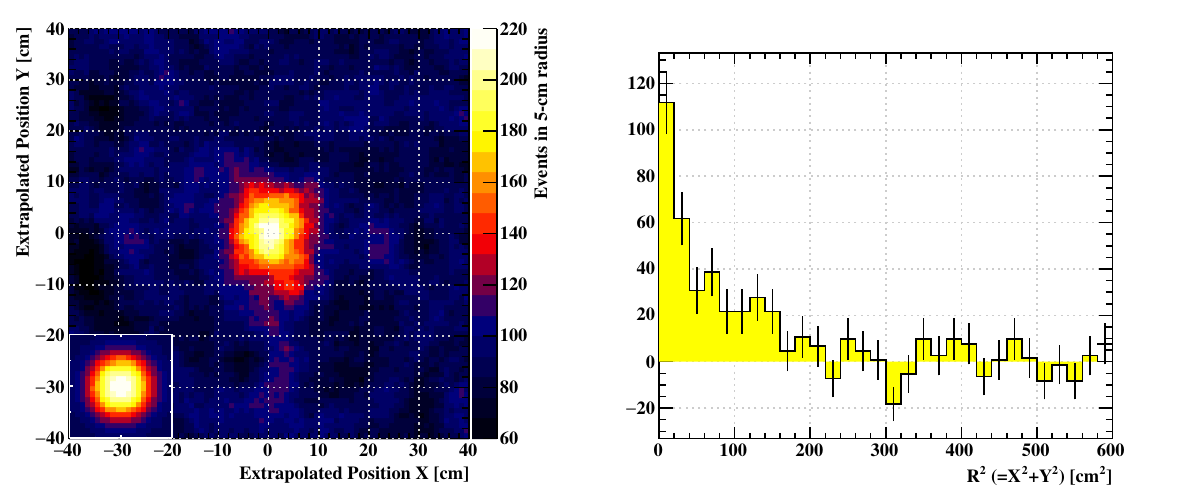}
\caption{
 Extrapolated position distribution (left) as a result of the counting of events with running bins (5-cm-radius circles). The energy range is set as 100--300 MeV. The inset shows the simulated point-source image in the energy region considered. Distribution of $R^2$ after the subtraction of the background contribution estimated by counting in off-source regions (right).
 }
\label{image}       
\end{figure}

Figure \ref{image} shows the distributions of the extrapolated positions in the energy range of 100--300 MeV. 
The left image shows the result of the counting the number of events smeared with a circle of a 5 cm radius; the radius is sufficiently small compared with the expectation by angular resolution (0.017 rad $\times$ 440 cm  = 7.5 cm at 100 MeV).
The simulated point-source image in the energy range of 100--300 MeV is also shown in the bottom-left inset. 
The right figure shows the distribution of $R^2$, where $R$ is the distance in the extrapolated position space between each gamma-ray event and the center of the detected image.
The background distribution, which can be predicted by counting in off-source regions, is subtracted from the entries in each bin.
 A clear excess in the number of events was observed, together with an extended structure, which is expected because of the size of the launching plate. These results demonstrate that we succeeded in the imaging of the launching plate by using gamma rays. 
\par

\subsection{Discussion}
\subsubsection{MC simulation}
To reproduce gamma rays emitted from the region of the launching plate, we performed the following MC simulation.
First, we simulated the targets that imitated shapes, materials and masses of the launching plate (4.6 kg, aluminum) and shackles (3.1 $\times$ 2 kg, stainless steel);
second, we placed them 4.4 m above the detector in the geometry of the Geant4 simulation.
Incident particles were generated at a plane over the targets. 
To estimate the energy spectra, zenith-angle distributions, and azimuthal-angle distributions of cosmic rays at the observation altitudes in Australia, we used the calculation results obtained from the HKKM model \cite{Honda}.
The results obtained were applied to the incident particles (protons, alpha particles, electrons, positrons, and gamma rays). The QGSP BERT package was used for the physical process model (PhysicsList). If the gamma rays reached the detector surface, then the information about those gamma rays produced by interactions were provided as output.

\subsubsection{Flux measurement}
\begin{figure}[]
\centering
\includegraphics[ width=12.0 cm,clip]{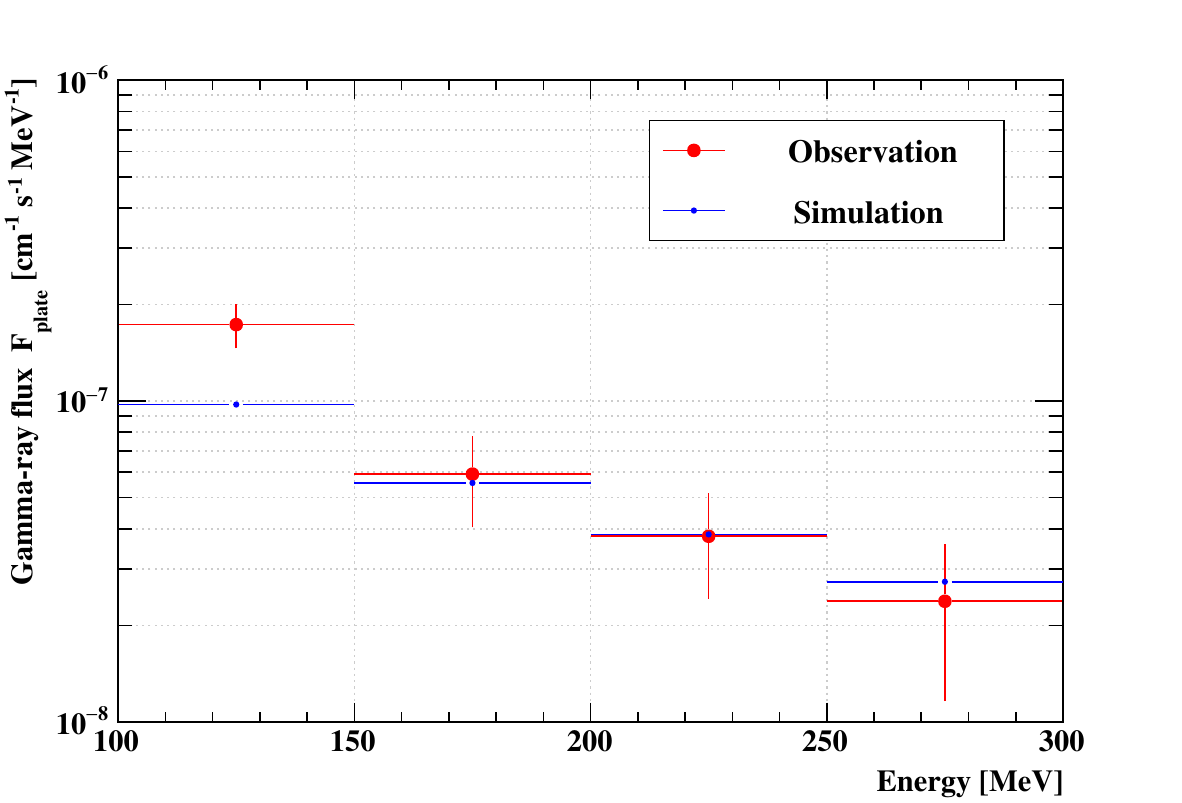}
\caption{
Gamma-ray flux from the launching-plate region at the position of the emulsion telescope.
}
\label{FluxAtDetector}       
\end{figure}

We defined $F_{plate}$ as the flux of gamma rays from the launching plate.
Signal events in the on-source region were set at $R<20$ cm.
Figure \ref{FluxAtDetector} shows the results of the measured (red points) and simulated (blue points) distributions for $F_{plate}$.
The emulsion telescope employed in the GRAINE 2015  experiment observed gamma rays from the launching plate with the expected flux.

\subsubsection{Imaging resolution}
\begin{figure}[]
\centering
\includegraphics[width=12.0 cm,clip]{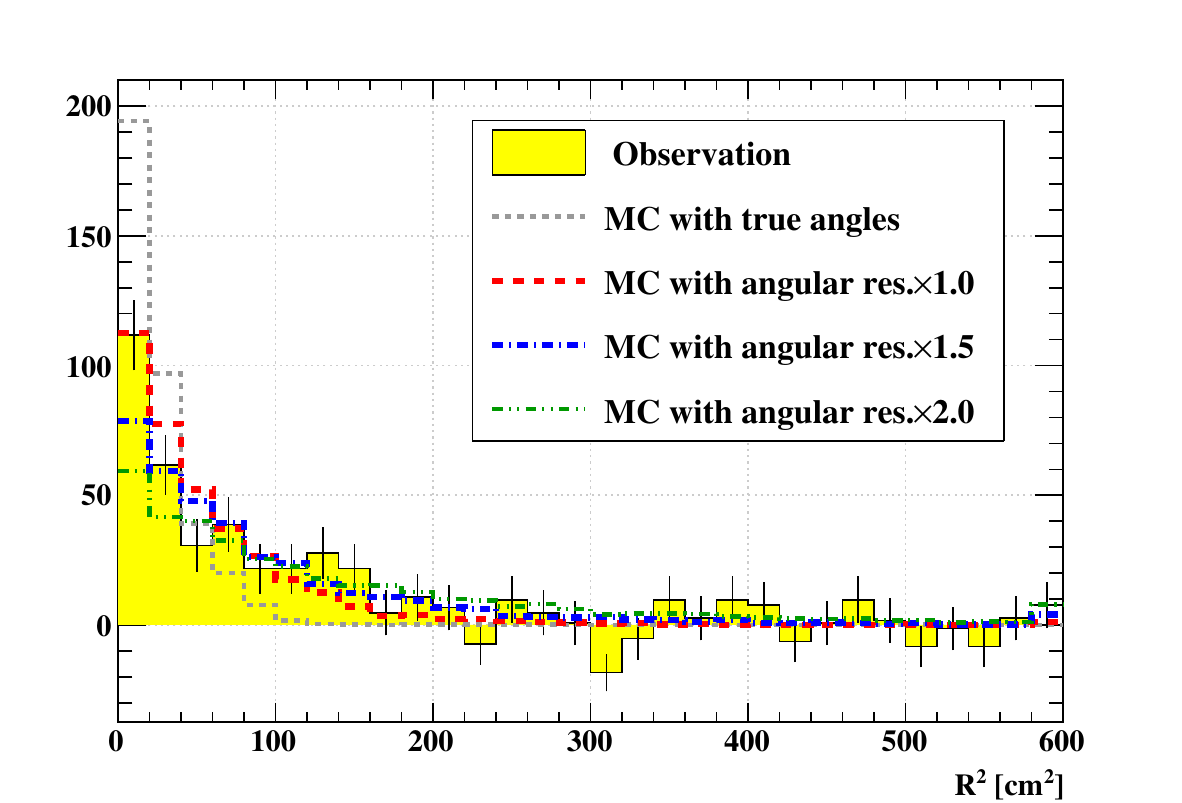}
\caption{
Experimental and MC-simulated $R^2$ distributions. }
\label{R2dist}       
\end{figure}

Figure \ref{R2dist} shows  experimental and simulated $R^2$ distributions.
The gray-dashed line indicates the simulated distribution with no angular errors on the gamma ray direction.
\par
The angular resolution of the emulsion telescope can be approximated by using the following equation: 
\begin{equation}
\theta_{68\%} \simeq \frac{100}{E_{\gamma}} \rm{[^{\circ}]}
\label{eq_resolution}
\end{equation}
where $\theta_{68\%}$ is the 68\% containment angle and $E_{\gamma}$ is the gamma-ray energy in MeV.
The red, blue, and green lines  indicate simulated distributions obtained by smearing the gamma-ray angles with angular resolutions $\theta_{68\%} \times a$, with $a=1.0$, 1.5, and 2.0, respectively.
Here, a simple 2-D Gaussian was used as point-spread function (PSF).
Reduced chi-squares (and p-values) for each value of $a$  are: $\chi^2_{1.0}/\rm{ndf} =1.67$ ($P_{1.0}=0.154$), $\chi^2_{1.5}/\rm{ndf}=2.27$ ($P_{1.5}=0.059$), and $\chi^2_{2.0}/\rm{ndf}=4.84$ ($P_{2.0}=0.001$)  in the region $R^2<100$ cm$^2$ where 
the differences among the 3 MC calculations mostly appear and the number of events in the observed data are statistically significant.
The red line ($a=1.0$) is the distribution that well reproduces the experimental data.
Therefore, we conclude that the emulsion telescope had an angular resolution that satisfied our expectations  in the energy region considered.

\section{Conclusion} 
We are promoting the gamma-ray observation project with excellent imaging, GRAINE, with balloon-borne emulsion gamma-ray telescopes that realize observations characterized by high angular resolution, polarization sensitivity, and large aperture area. 
We conducted a balloon experiment in 2015, by performing also data acquisition of the recovered emulsion films.
We realized stable data acquisition by using the latest emulsion scanning system, HTS.
In this way, owing to the introduction of a highly sensitive film, we obtained a high S/N ratio and a high track-finding efficiency data. Emulsion films with a total area of 41 m$^2$ have been scanned in three months, which is the fastest data-acquisition period reported thus far in the context of emulsion experiments. 
\par
We established an automatic selection for electron-pair events.
The selection of conversion points, TFD, and the energy reconstruction were performed systematically.
In the GRAINE 2015 gamma-ray analysis, 
the fraction of the total background events was evaluated as 5.3\%.
The selection efficiencies for 100- and 200-MeV gamma-rays were 65\% and 83\%, respectively, according to MC simulation, which are approximately doubled compared with the previous experiment (GRAINE 2011).
The atmospheric gamma-ray flux at the balloon altitude derived from the observation data is consistent with the one obtained in previous measurements.
By automatizing the selection process, we analyzed almost the whole converter part and recorded approximately $10^6$ gamma-ray events. 
\par		
We conducted the demonstration analysis of gamma-ray imaging of the emulsion telescope and succeeded in the detection of an external gamma-ray source, i.e., the launching plate interacting with cosmic rays.
The measured flux from the source region and the PSF of the image were consistent with those obtained from MC simulation.
In particular, we confirmed that the emulsion telescope had an excellent angular resolution in the 100--300-MeV energy region during the observation, as expected.
We previously demonstrated the imaging performance with a small area, by performing beam tests with an accelerator for gamma rays.
However, the results reported in this paper correspond to the first analysis of the entire area of an enlarged detector using an external gamma-ray source at the balloon altitude, and indicate an important milestone for the development of large-area emulsion gamma-ray telescopes. 
\par
The results of the fast data acquisition, the establishment of an automatic selection analysis, and the demonstration of the imaging performance with the whole chamber, as previously mentioned, have strongly proven that balloon-borne experiments with a larger-area emulsion telescope are feasible. 
\par
We are preparing for the next balloon experiment, which is planned for April 2018, and the instruments have been further updated to increase the yields of the gamma rays (e.g., we have constructed a star camera system to determine the attitudes with a higher efficiency).
The goal of this experiment is to detect celestial gamma-ray sources. After the final confirmation of the overall performance of the balloon-borne emulsion telescope, the GRAINE project will enter into the scientific observation phase. 

\section*{Acknowledgment}
The scientific balloon (DAIKIKYU) flight opportunity was provided by ISAS/JAXA. 
We would like to thank Dr. M. Honda (ICRR, University of Tokyo) for MC calculations of the cosmic-ray flux at balloon altitudes.
This work was supported by JSPS. KAKENHI (Grant Numbers 26247039, 26105510, 26800138, and 16K17691) and a Grant-in-Aid for JSPS. Fellows (13J09408).


\begin{thebibliography}{9}
%
\bibitem{AGILE} M. Tavani et al., Astron. Astrophys. 502, 995 (2009).
\bibitem{LAT} W. B. Atwood et al., Astrophys. J. 697, 1071 (2009).
\bibitem{catalog} F. Acero, et al., Astrophys. J. Suppl. Ser. 218 (2), 23, (2015).
\bibitem{cr} A. Giuliani, et al., Astrophys. J. Lett. 742 (2), L30, (2011).
\bibitem{crfermi} M. Ackermann, et al., Science 339 (6121), 807-811, (2013).
\bibitem{GeVexcess} T. Daylan, et al. Physics of the Dark Universe 12 ,1-23, (2016).
\bibitem{J.Takata} J. Takata and H.K. Chang, Astrophys. J. 670, 677 (2007).
\bibitem{INTEGRAL}A.J.Dean, et al.,Science, 321, 1183 (2008).
\bibitem{INTEGRAL2} M. Forot, et al., Astrophys. J. 668, 1259 (2007).
\bibitem{GAP} D. Yonetoku, et al.  Astrophys. J Letters 743.2, L30 (2011).
\bibitem{PoGo} M. Chauvin, et al., Scientific Reports 7 (2017).
\bibitem{nakano} K. Morishima and T. Nakano, J. Instrum. 5, P04011 (2010).
\bibitem{polarization} K. Ozaki, et al., Nucl. Instrum. Meth. A 833,165-168,  (2016).
\bibitem{shifter} S. Takahashi et al., Nucl. Instrum. Meth. A 620, 192 (2010).
\bibitem{JACEE} T. Burnett, et al., Nucl. Instrum. Meth. A 251, 583  (1986).
\bibitem{RUNJOB} A.V. Apanasenko, et al., Astropart. Phys. 16, 13 (1986) .
\bibitem{COSPAR} S. Takahashi, et al., Advances in Space Research, DOI: 10.1016/j.asr.2017.08.029
\bibitem{Rokujo} H. Rokujo et al., Nucl. Instrum. Meth. A 701, 127 (2013).
\bibitem{GRAINE2011} S. Takahashi et al., Prog. Theor. Exp. Phys. 2015, 043H01 (2015).
\bibitem{GRAINE2015} S. Takahashi et al., Prog. Theor. Exp. Phys. 2016, 073F01 (2016).
\bibitem{Mizutani} F. Mizutani et al., Nucl. Instrum. Meth. A, submitted.
\bibitem{gondola} H. Rokujo et al., Proceedings of Science (ICRC2015), 1021, (2017)
\bibitem{Camera} K. Ozaki et al., Proc. Balloon Symp., ISAS/JAXA, isas15-sbs-034 (2015) (in Japanese).
\bibitem{ozaki} K. Ozaki et al., J. Instrum. 10, P12018 (2015).
\bibitem{HTS} M. Yoshimoto et al., Prog. Theor. Exp. Phys. 2015, 103H01 (2015).
\bibitem{taku} T. Nakamura et al., Nucl. Instrum. Meth. A 556, 80 (2006).
\bibitem{NETSCAN} K. Kodama et al., Nucl. Instrum. Meth. A 493, 45 (2002).
\bibitem{hamada} K. Hamada et al., J. Instrum. 7, P07001 (2012).
\bibitem{Fukuda} T. Fukuda et al.,  J. Instrum. 5, P04009 (2010).
\bibitem{Park} K. Kodama et al., Nucl. Instrum. Meth. A 574, 192 (2007).
\bibitem{Geant} S. Agostinelli et al., Nucl. Instrum. Meth. A 506, 250 (2003).
\bibitem{Kawahara} H. Kawahara et al., Proceedings of Science (KMI2017), Vol.294 059, (2017)
\bibitem{Honda} M. Honda et al., Phys. Rev. D 92, 023004 (2015).



\end{thebibliography}
\end{document}